# Designing a Dynamic Components and Agent based Approach for Semantic Information Retrieval

Zeeshan Ahmed, Detlef Gerhard

Mechanical Engineering Informatics and Virtual Product Development Division (MIVP),
Vienna University of Technology,
Getreidemarkt 9/307 1060 Vienna, Austria
{zeeshan.ahmed, detlef.gerhard} @tuwien.ac.at

*Abstract*— **In this paper based on agent and semantic web technologies we propose an approach .i.e., Semantic Oriented Agent Based Search (SOAS), to cope with currently existing challenges of Meta data extraction, modeling and information retrieval over the web. SOAS is designed by keeping four major requirements .i.e., Automatic user request handling, Dynamic unstructured full text reading, Analysing and modeling, Semantic query generation and optimized result classifier. The architecture of SOAS is consisting of an agent called Personal Agent (PA) and five dynamic components .i.e., Request Processing Unit (RPU), Agent Locator (AL), Agent Communicator (AC), List Builder (LB) and Result Generator (RG). Furthermore, in this paper we briefly discuss Semantic Web and some already existing in time proposed and implemented semantic based approaches.**

## I. INTRODUCTION

D RIVEN by problems raised in web technology technologic developments [17], especially publishing data in machine read and process able format to take advantage in knowledge based information extraction and maintenance [1]. Currently it is not possible to search and extract optimized results using full text queries because there is no such mechanism exists which can fully extract the semantic from full text queries and then look for particular knowledge based information or publish data over the web in meaning full structured way [5]. These problems require a comprehensive solution to read, understand and convert full text unstructured data in to Meta data [6]. A number of semantic based solutions have been developed, including Semantic Desktop [10], Reisewissen [11], Test Application [15] and Metadata Search Layer [16] but still the problems are not solved.

Mechanism of presenting information over the web in a format so that human being as well as machines can understand the context leads to the concept of Semantic Web introduced by Tim Berners Lee [7]. Semantic web is a linked mesh of information to produce technologies capable of reasoning on semi structured information [9] and process by machines [1]. Currently, semantic web is standing on two very important building blocks Ontology and Semantic Web Agent [3].

Ontology is the explicit representation and description of already available finite sets of terms and concepts used to make the abstract model of a particular domain. It includes information in the form of properties containing values, restrictions and relationship with the other properties. Relationships are based on classes and their sub classes which can be implemented using different Ontology supported languages like Resource Description Framework (RDF) [1] and Web Ontology Language (OWL) [8]. Semantic web agent is an advanced implemented form of object oriented concepts, having autonomous behavior and use the concepts of metadata in solving problems of a specific domain. Moreover, along with the processing ability semantic web agent is also capable of communicating, receiving and transferring information to different sources over the web.

We briefly address some already proposed and implemented methodologies providing values to semantic web problems in section 2. We propose an approach to provide comprehensive support in implementing a semantic and agent based solution towards the problems of Meta data extraction, modeling and information retrieval over the web in section 3.

## II. RELATED RESEARCH WORK

### A. Semantic Desktop [10]

Authors proposed an approach, promoting the idea of stepping into user's mental model by implementing a Personal Information Model (PIM). PIM is designed to improve the process for the identification of documents and retrieval of no unnecessary document. The proposed architecture mainly consists of three main components Receiver, Interpreter and Analyzer. Receiver provide index services to obtain the information about files structure, Interpreter structure and model the contents of obtained information and Analyzer query created models to infer, run and integrate rules. Authors successfully tested the proposed approach in designed four case scenarios local search, group search, closed community and open community.

### B. Reisewissen [11]

Authors proposed an approach to identify the potential relevant knowledge sources and provide quality services by semantically connecting, organizing and sharing the currently isolated pieces of information in an online portal to anticipate customer behavior. Proposed approach is implemented using semantic web technologies in a project Reisewissen, a hotel recommendation engine and travel information system. The design architecture of Reisewissen consists of three main components Data Connectors (DC), Evaluation Framework (EF) and Evaluation Engine (EE). DC provide transparency to data sources, transformation of data from heterogeneous to common data format and caching of data, EF test the quality of data and EE rank and filter information by their weights.

### C. Test Application [15]

Authors proposed an approach based on the semantic web concepts for publishing information on web without inserting into relational database by making a flexible reasoning Ontology based system and to take advantage in improving already existing semantic search mechanism. According to the proposed architecture, each web object is referred as Unified Resource Identifier (URI) and context is provide in machine interpretable way. The proposed approach is successfully implemented in real time software application called Test Application, to publish the information in to machine interpretable form for mobile phones.

### D. MetaData Search Layer [16]

Authors proposed a successful process for metadata search to identify the location from set of locations contained by a document and avoid looking into non-specific document. Scalability and efficiency of approach is determined using simulation of document metadata keywords, location pointers, node connections and node knowledge.

## III. SEMANTIC ORIENTED AGENT BASED SEARCH

We propose Semantic Oriented Agent based Search (SOAS), dynamic components and agents based approach to handle user's unstructured full text requests by converting in to structured information models and generating semantic based queries to search particular information. SOAS is designed by keeping four major requirements in mind which are; automatic user request handling, dynamic unstructured full text data reading, analyzing and modeling, and semantic query generation and optimized result classifier.

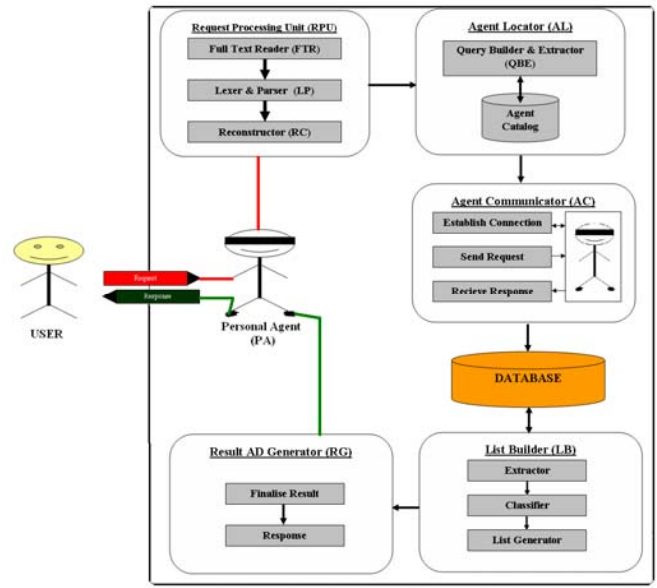

Fig1. The architecture of Semantic Oriented Agent based Search [18]

The proposed designed architecture of SOAS, as shown in Fig 1, is consist of one Personal Agent (PA), five dynamic processing units .i.e., Request Processing Unit (RPU), Agent Locator (AL), Agent Communicator (AC), List Builder (LB) and Result Generator (RG) and Database.

### A. Personal Agent (PA)

PA is an agent based software application having autonomous behavior for working in a particular domain by communicating and collaborating with some other software applications in an automated environment to obtain desired results. The job of PA in SOAS is to communicate with user, take user's web based search requests in unstructured full text form and forward these requests for further processing and responds back to the user with obtained results.

*B. Request Processing Unit (RPU),*

RPU is the most important component of the SOAS. The job of RPU is to get data request in an unstructured full text form, structure the contents, extract the semantic and reconstruct a structured semantic based data request. To fulfill the job requirements, RPU is divided into three sub components .i.e., Full Text Reader (FTR), Lexer & Parser (LP) and Reconstructor (RC). The job of FTR is quite simple; it is required to just receive unstructured full text requests and forward one request at a time to LP. LP is the most important sub component of the RPU, its job is to take unstructured full text, generate lexical tokens, apply natural language parser rules for semantic analysis based on used natural language grammar like English, filters recognized and unrecognized contents of request and then passed only semantically recognized and verified contents to RC, the last subcomponent of RPU. The job of RC is to receive semantically verified contents, using LiveLink, a database driven web based Knowledge Management system [12], structure the contents and create Ontologies based on RDF models.

*C. Agent Locator (AL)*

AL is designed to search and extract the contact information of particular agent of a particular domain using forward changing principle, first look for particular domain and then try to find out the contacts of available particular agents of that domain. AL is consist of a sub component Query Builder & Extractor (QBE) and database Agent Catalog, QBE use RDF models and create search query to extract contact information of available agents from Agent Catalog.

*D. Agent Communicator (AC),*

The job of AC is to get the information based on user request using the contacts of available agents extracted and input by AL. AC is divided in to three sub components to establish connection with agents, communicate with connected agents by sending requests and receiving responses. Furthermore the obtained information from respective agents during communication is stored in database.

*E. List Builder (LB)*

The job of LB is to retrieve stored results from database, if results are more than one then rank them and create a prioritized list of results. To full fill the desired job LB is divided in to three sub components .i.e., Extractor, Classifier and List Generator. Extractor is designed to extract stored results from database, Classier is used to classify the result on the basis of their weights and List Generator create list of prioritized results.

*F. Result Generator (RG)*

The job of RG is to convert finalized list of prioritized into the acceptable readable format. RG is consisting of two sub components Finalise Result and Response. Finalise Result convert results input by LB in acceptable readable format and then Response simply forward these results to PA.

The architecture is designed in a way that the unstructured full text user request desiring some particular knowledge based information is given to PA. PA will simply transfer the received request to RPU, which will analyze and convert unstructured full text user request in to a structured semantic based data request. Constructed new semantic based search request will be passed and used by AL to find out the contact information of particular domain's available agents. Then using extracted contacts, AC will contact, communicate and obtain information from available agents, and stored in database. Stored information will be extracted and prioritized list of results will be generated by LB. Prioritized results will be forwarded to RG, which will finalize the result by converting into the acceptable readable format and pass to PA. In the end PA will respond back to user by providing retrieved results.

IV. CONCLUSION

In this paper we have briefly discussed the semantic web technologies, Meta data extraction, modeling and information retrieval over the web problem as the most important of all existing web problems. We have introduced our own approach Semantic Oriented Agent based Search (SOAS), a dynamic components and agent based approach, as the solutions and briefly presented the designed architecture. In future, to evaluate the effectiveness of SOAS, we will implement it in a real time software application using already existing desktop, web and semantic web technologies.

ACKNOWLEDGEMENT

The authors would like to acknowledge the support of Mechanical Engineering Informatics and Virtual Product Development Division (MIVP), Vienna University of Technology Austria.

REFERENCES

[1] Sean B. Palmer, "The Semantic Web: An Introduction", viewed February, 28. 2007, <http://infomesh.net/2001/swintro>

[2] Ora Lassila and Ralph R. Swick, "Resource Description Framework (RDF) Model and Syntax Specification", viewed February, 28. 2007, <http://www.w3.org/TR/1999/REC-rdf-syntax-19990222>

[3] Wernher Behrendt, "Ambient Intelligence Semantic Web or Web 2.0", *In Proceedings of Semantic Content Engineering 2005*, Salzburg 2005

[4] What Is An RDF Triple? , viewed February, 28. 2007, <http://www.robertprice.co.uk/robblog/archive/2004/10/What_Is_An_RDF_Triple_.shtml>

[5] Tim Berners-Lee, James Hendler and Ora Lassila, "The Semantic Web, A new form of Web content that is meaningful to computers will unleash a revolution of new possibilities", May 2001, Reviewed February 2007, <http://www.sciam.com/article.cfm?articleID=00048144-10D2-1C70-84A9809EC588EF21>

[6] Danbri, "Metadata and Resource Description", 5 April 2001, viewed February, 28. 2007, <http://www.w3.org/Metadata>

[7] Tim Berners-Lee, 05 December 2006, viewed February, 28. 2007, <http://www.w3.org/People/Berners-Lee>

[8] OWL Web Ontology Language, viewed February, 28. 2007, <http://www.w3.org/TR/owl-features>

[9] Sebastian Ryszard Kruck, Mariusz Cygan, Piotr Piotrowski, Krystian Samp, Adam Westerski and Stefan Decker, "Building a Heterogeneous Network of Digital Libraries on Semantic Web", *In Proceedings of Semantic Systems From Visions to Applications*, Vienna Austria 2006

[10] Mark Siebert, Pierre Smits, Leo Sauermann and Andreas Dengel, "Increasing Search Quality with the Semantic Desktop in Proposal Development", *In the proceedings of Practical Aspects of Knowledge Management 6th International Conference*, PAKM , Vienna Austria 2006

[11] Magnus Niemann, Malgorzata Mochol and Robert Tolksdorf, "Improving Online Hotel Search – What Do We Need Semantic For", *In Proceedings of Semantic Systems from Visions to Applications*, Vienna Austria 2006

[12] Livelink, viewed February, 28. 2007, <http://www.greggriffiths.org/livelink>

[13] Jena – A Semantic Web Framework for Java, viewed February, 28. 2007, <http://jena.sourceforge.net>

[14] Guizhen Yang and Michael Kifer, "Reasoning about Anonymous Resources and Meta Statements on the Semantic Web", Journal on Data Semantics (JoDS), Volume I, pp. 69-97, 2003

[15] Markus Linder, Martin Schliefnig, Dieter Fensel, Schahram Dustdar, Heinrich Otruba and Tassilo Pellegrini, "The realization of Semantic Web based E-Commerce and its impact on Business, Consumers and the Economy", *In Proceedings of Semantic Systems From Visions to Applications*, Vienna Austria 2006

[16] Sam Joseph, "P2P MetaData Search Layers", *In the proceedings of Agents and Peer-to-Peer Computing, Second International Workshop*, AP2PC 2003, Melbourne Australia 2003

[17] World Wide Web, March 2007, <http://en.wikipedia.org/wiki/World_Wide_Web>

[18] Zeeshan Ahmed, Detlef Gerhard, "An Agent based Approach towards MetaData Extraction and Information Retrieval over the Web", Accepted at First International Workshop on Cultural Heritage on the Semantic Web in conjunction with the 6th International Semantic Web Conference and the 2nd Asian Semantic Web Conference, 2007, Busan,Korea November 2007